\documentclass[aps,prl,twocolumn,showpacs,floatfix]{revtex4}
\usepackage{graphicx,color}
\usepackage{amsmath,amssymb}
\begin{document}
\title{A charged particle in a magnetic field - Jarzynski Equality}
\author{A.M. Jayannavar }
\email{jayan@iopb.res.in }
\author{Mamata Sahoo}
\affiliation{Institute of Physics, Bhubaneswar 751005 India.}
\date{\today}
\begin{abstract}
Abstract:~~ We describe some solvable models which illustrate the Jarzynski theorem and related fluctuation theorems.~We consider a charged particle in the presence of magnetic field in a two dimensional harmonic well.~In the first case the centre of the harmonic potential is translated  with a uniform velocity,~while in the other case the particle is subjected to an ac force.~We show that Jarzynski identity  complements Bohr-van Leeuwen theorem on the absence of diamagnetism in equilibrium classical system.
\end{abstract}
\maketitle

Most processes that occur in nature are far from equilibrium and hence cannot be treated within the framework of classical thermodynamics.~The traditional nonequilibrium statistical mechanics deals with systems near equilibrium in the linear response regime.~Its success has lead to the formulation of fluctuation-dissipation relation,~Onsagar's reciprocity relations and the ~Kubo-Green formulae,~ etc.~~However,~very recent developments in nonequilibrium statistical mechanics have resulted in  the discovery of some exact theoretical results for systems driven far away from equilibrium and are collectively called fluctuation theorems[1].~These results include entropy production theorems[2],~Jarzynski equality[3],~Crooks relations[4],~Hatano-Sasa identity[5],~etc.~Some of the above relations have been verified experimentally on single nanosize systems in physical envirnoments where fluctuations play a dominant role[6,7].

The concept of free energy is of central importance in statistical mechanics and thermodynamics.~With the help of free energy one can calculate all the phases of a system and their physical properties.~~However,~the free energy of the system relative to an arbitrary reference state is often difficult to determine.~Jarzynski equality(JE) relates non-equilibrium quantities with equilibrium free energies[3].~~ In this prescription,~ initially the system is assumed to be in equilibrium state determined by a thermodynamic parameter~$A$~ defined by a control parameter $\lambda_{A}$ and is kept in contact with a heat bath at temperature T.~~The nonequilibrium process is obtained by changing the thermodynamic control parameter  $\lambda$ in a finite time $\tau$ according to a prescribed protocol $\lambda(t)$,~from $\lambda_{A}=\lambda($t=0$)$ to some final value $\lambda_{B}=\lambda(t=\tau)$.~~The final state of the system at time $\tau$ (at the end of the protocol) will in general not be at equilibrium.~It will equilibrate to a final state $B(\equiv \lambda_{B})$ if it is further allowed to evolve by keeping parameter $\lambda_{B}$ fixed.~~JE states that
\begin{equation}
\langle \exp(\frac{-W}{k_{B}T}) \rangle=\exp(\frac{-\Delta F}{k_{B}T})
\end{equation} 
Where $\Delta F$ is the free energy difference between equilibrium states $A$ and $B$.~ Angular bracket $\langle... \rangle$ denotes the average taken over different realizations for fixed protocol $\lambda(t)$.~~$W$ is work expended during each repetation of the protocol and is a realization dependent random variable.~Jarzynski's theorem has been derived using various methods with different system dynamics[3-5,8].~This remarkable identity provides a practical tool to determine equilibrium thermodynamic potentials from processes carried out arbitrarily far away from  equilibrium.~This identity has been used to extract equilibrium free energy differences in experiments.~Work distributions have been calculated analytically for several model systems and tested against various fluctuation theorems[9-11].~JE has been generalized to arbitrary transitions between nonequilibrium steady states by Hatano and Sasa[5],~which has also been verified experimentally[7].

In our present work,~we determine the work distributions analytically for two different models.~In both these cases charged particle dynamics in a two dimensional harmonic trap in the presence of a magnetic field is considered.~In the first case (i),~centre of the harmonic trap is dragged with a uniform velocity whereas in the case (ii),~ the particle is subjected to an ac force.~~We show that JE is consistent with  Bohr-van Leeuwen theorem.~ We also discuss steady state fluctuation theorems [SSFT] and energy loss in driven systems.

We consider a charged particle motion in two dimensions (x-y plane) in the presence of a time dependant potential $U(\equiv U(x,y,t))$.~~An external magnetic field $B$ along z-direction will produce Lorentz force on the charged particle.~The interaction of the particle with   environment can be treated via frictional force along with concomitant fluctuations.~~The appropriate equations of motion are given by the Langevin's equations[12,13]
\begin{equation}
m\ddot{x} =-\gamma \dot{x}-\frac{\vert{e}\vert}{c}B\dot{y}-\frac{\partial U}{\partial x}+\xi_x(t),
\end{equation}
 \begin{equation}
m \ddot{y}=-\gamma \dot{y}+\frac{\vert{e}\vert}{c}B \dot{x}-\frac{\partial U}{\partial y}+\xi_y(t),
\end{equation}
where the random force field $\xi_{\alpha}(t)$ is a Gaussian white noise.i.e,
\begin{equation}
\langle \xi_{\alpha}(t)\xi_{\beta}(t^{'}) \rangle=D\delta_{\alpha \beta} \delta(t-t^{'}) ,
\end{equation}
with $\alpha,\beta =x,y$,~e is the charge of the particle.~  $D=2\gamma k_{B} T$~~ is a consistency
condition for the system to approach equilibrium in the absence of time independent potential .~The friction coefficient is denoted by $\gamma$.

This problem (for time independent potentials) was earlier considered[13] to elucidate the crucial role played by the boundary conditions in the celebrated theorem of Bohr-van Leeuwen on the absence of diamagnetism in classical systems[13,14].~~This theorem states   that in equilibrium for classical systems the free energy is independent of magnetic field.~  Hence,~  diamagnetism does not exist in classical statistical mechanics.

We  restrict our analysis  to overdamped regime where the corresponding dynamical equations become
\begin{equation}
\gamma \dot{x}=-\frac{\vert e \vert B}{c}\dot{y}-\frac{\partial U}{\partial x}+\xi_x(t)
\end{equation}

\begin{equation}
\gamma \dot{y}=\frac{\vert e \vert B}{c}\dot{x}-\frac{\partial U}{\partial y}+\xi_y(t)
\end{equation}
  The associated Fokker-Planck equation[15]  leads to an equilibrium  distribution ~$P_{e}\propto e^{-(\frac{U(x,y)}{k_{B}T})}$ ~in a stationary regime for time independent potential $U(x,y)$.~This  distribution $P_{e}$ is independent of magnetic field consistent with Bohr-van Leeuwen theorem.

We consider in this work two cases for time dependent potentials.~For case(i) $U(x,y,t)=\frac{1}{2} k \vert \vec{r}-\vec{r}^{*}(t) \vert^{2}$,~where
$\hat{r}$ is a two dimensional vector $\vec{r}=\hat{i}x+\hat{j}y$ and 
$\vec{r}^{*}(t)=vt(\hat{i}+\hat{j})$,~$\hat{i}$ and $\hat{j}$ are unit vectors along $x$ and $y$ direction respectively.~Here the  particle is in a harmonic potential whose centre is dragged along with an uniform speed $\sqrt{2}v$ in a diagonal direction.~This problem can also be solved  for the  motion of the centre in an arbitrary direction with a different protocol.~For case(ii),~$U(x,y,t)=\frac{1}{2} k (x^{2}+y^{2})-A x \sin(\omega t)$ .~Here the particle in a two dimensional harmonic well is subjected to an ac force in x-direction.~~This problem  can also be solved for ac drivings in both $x$ and $y$ direction with different amplitudes and with a phase difference.

We rewrite equations (5) and (6) using the variable[12] $z=x+iy$,($i=\sqrt{-1}$).~For the case(i) we get 
\begin{equation}
\dot{z}=\frac{-kpz}{\gamma}+\frac{kpg^{*}(t)}{\gamma}+\frac{p\xi(t)}{\gamma} ,
\end{equation}
where $p=\frac{1+iC}{1+C^{2}}$, $\xi(t)=\xi_{x}(t)+i\xi_{y}(t)$,~   
$g^{*}(t)=vt(1+i)$ and $C=\frac{e \vert B \vert}{\gamma c}$\\
For the case(ii),~we get
\begin{equation}
\dot{z}=\frac{-kpz}{\gamma}+\frac{p}{\gamma}(A\sin(\omega t)+\xi(t)) .
\end{equation}
Thermodynamic work $W$ done on the system by an external agent during a time interval $t$ is given by[9,10],~
for case(i),
\begin{equation}
W=-kv\int_0^{t} \{(x-vt^{'})+(y-vt^{'})\} dt^{'} ,
\end{equation}
and for case(ii)
\begin{equation}
 W=-A\omega \int_0^t \cos(\omega t^{'})  x(t^{'}) dt ,
\end{equation}
The formal solution of equations (7) and (8) respectively are 
 \begin{eqnarray}
z(t)&=&z_{0} \exp(-\frac{k}{\gamma}pt)+\frac{p}{\gamma} \int_0^t dt^{'} \exp(-\frac{k}{\gamma}p(t-t^{'}))\{kg^{*}(t^{'})\nonumber\\&&+\xi(t^{'})\} ,
\end{eqnarray}
and
\begin{eqnarray}
z(t)&=&z_{0} \exp(-\frac{k}{\gamma}pt)+\frac{p}{\gamma} \int_0^t dt^{'} \exp(-\frac{k}{\gamma}p(t-t^{'}))\{\xi(t^{'})\nonumber\\&&+A\sin(\omega t^{'}))\} .
\end{eqnarray}
where $z_{0}=x_{0}+iy_{0}$,~~$x_{0}$ and $y_{0}$ are initial co-ordinates
of the particle.~The initial distribution for $x_{0}$ and $y_{0}$ is assumed to be equilibrium canonical distribution $P_{e}(x_{0},y_{0},t)=\frac{\beta k}{2\pi} \exp[\frac{-\beta k (x_{0}^{2}+y_{0}^{2})}{2}]$.~~It may be readily noted from equations (10),(11),(12) and (13) that work done as well as particle co-ordinates at later times are linear functional of Gaussian variables $\xi_{x}(t)$ and $\xi_{y}(t)$ and hence their distributions are Gaussian.~We calculate full probability distribution for
$W$ for both cases following closely the procedures adpoted in refs[9,10] .~Without going into further details of algebra we give our final results which will be further analyzed.

For the case(i),~the average work done $\langle W \rangle$ is
\begin{eqnarray}
\langle W \rangle &=& 2 \gamma v^2 \{t-\frac{\gamma}{k}(1-\exp(-k^* t)
\cos(\Omega t))-\frac{C \gamma}{k}\sin(\Omega t)\nonumber\\&&\exp (-k^* t)\}-\gamma v^2 2 C\{\frac{\gamma}{k}\sin(\Omega t)\exp(-k^*t)-\frac{C \gamma}{k}\nonumber\\&&(1-\exp(-k^*t)\cos(\Omega t))\} ,
\end{eqnarray}
where
$\Omega = \frac {k C}{\gamma(1+C^2)}$ and 
$k^{*}=\frac{k}{\gamma(1+C^{2})}$.~The above equation(13) agrees with the result obtained in refs[9,10] for $B=0$.~The variance of the work is given by
\begin{equation}
\langle W^2 \rangle -\langle W \rangle^2 = \frac{2 \langle W \rangle }{\beta} ,
\end{equation}
where $\beta=\frac{1}{k_B T}$.~the full probability distribution $P(W)$ is
\begin{eqnarray}
P(W)&=&\frac{1}{\sqrt{4\pi\langle W \rangle/\beta}} 
e^{-(W-\langle W \rangle)^{2}/(4\langle W \rangle/\beta)} .
\end{eqnarray}
 JE  given in equation(1) follows immediately from the above expression,~with 
\begin{equation}
\langle e^{-\beta W} \rangle= e^{-\beta \Delta F}=1 .
\end{equation}
The equation(16) implies $\Delta F=0$,~indicating that the equilibrium 
free energy of a particle in a harmonic potential is independent of magnetic field,~consistent with the Bohr-van Leeuwen theorem.~Needless to say that in the present case the free energy of the oscillator is also independent of position of the centre of the harmonic potential as expected on general grounds.~However,~it is interesting to note that the thermodynamic work $W$ in the transient state depends explicitly on the magnetic field $B$.~In the presence of magnetic field relaxation rate of the system  $\tau_{r}(=\frac{\gamma(1+C^{2})}{k})$ depends on the magnetic field.~It increases with the strength of the magnetic field.~Hence the magnetic field gives an additional control over the relaxation time in experimental situation to verify the above and subsequent results.

Only in the asymptotic time limit $t \rightarrow \infty$($t \gg \tau_{r}$),~where the system does not retain the memory of the initial state,~one obtains for the averaged work done $\langle W_{s} \rangle$ in a system in time interval $t$,
\begin{equation}
\langle W_{s} \rangle \approx 2 \gamma v^{2} t
\end{equation}
Hence power($p$) delivered to the system in the steady state is constant($p=2\gamma v^{2}$).~In this state thermodynamic work is essentially a mechanical work delivered to the system by a moving trap with a speed $\sqrt{2}v$ along the diagonal direction in x-y plane.~The particle in this state settles to a Gaussian distribution with the same dispersion as in the case of canonical equilibrium distribution.~However,~the centre of the position of the particle distribution lags behind the instantaneous minimum of the confining potential by a distance $l=\frac{\sqrt{2} v\gamma}{k}$ along the diagonal line.~The harmonic potential pulls the particle with a force  $kl$,~at speed $\sqrt{2} v$.~Thus the power delivered is $kl\sqrt{2} v =2 \gamma v^{2}$.~This power is dissipated as heat into the surrounding  medium.~~In this regime,~the fluctuations in  work
$\langle W_{s} \rangle$ is
\begin{equation}
\langle W_{s}^{2} \rangle = \langle W_{s} \rangle^{2}+\frac{2}{\beta}\langle W_{s} \rangle
\end{equation}
In the steady state distribution of work $W_{s}$ obeys the SSFT[9] namely 
\begin{equation}
\frac{P(W_{s})}{P(-W_{s})}=e^{\beta W_{s}}
\end{equation}

It is important to note that both work and its distribution in the steady state do not depend on the magnetic field.~Hence the experimentally obtained results[7] to verify Hatano-Sasa identity[5] for transition between nonequilibrium steady states,~ remains unaltered irrespective of the magnetic field being present or not . ~Here transition between nonequilibrium states are induced by varying the speed of the trap[7].

We now turn to case(ii).~~As mentioned earlier for this case the work distribution is Gaussian and is characterized completely by the first moment and the variance.~~Expression for $\langle W \rangle$ is too lengthy  and is given in appendix A.~The variance  is given by
\begin{equation}
\langle W^2 \rangle -\langle W \rangle^2 = \frac{2}{\beta}(\langle W \rangle -\Delta F)
\end{equation}
Where $\Delta F=\frac{-A^{2}}{2k} \sin^{2}(\omega \tau)$.~Using the distribution of work one can readily verify the JE,~namely 
$\langle e^{-\beta W} \rangle= e^{-\beta \Delta F}$.~$\Delta F$ is the free energy difference between two thermodynamic states.~ The initial ($t=0$) and final states ($t=\tau$) are characterized by a two dimensional harmonic potential with an additional tilt of magnitude zero and $-Ax\sin(\omega \tau)$, respectively.~Note that $\Delta F$ is independent of magnetic field,~again reassuring  Bohr-van Leeuwen theorem.~However,~  the work distribution is  an explicit function of magnetic field.

We now concentrate on the statistics of the work done $W_{s}$ in the asymptotic regime.~In this regime probability distributions are time periodic with a period $\frac{2\pi}{\omega}$.~We calculate averaged work done over one period $\frac{2\pi}{\omega}$,~which is
\begin{equation}
\langle W_{s} \rangle=\lim_{t \rightarrow \infty}[\langle W(t+\frac{2\pi}{\omega})\rangle-\langle W(t)\rangle]
\end{equation}

\begin{equation}
 \langle W_{s} \rangle=\frac{\pi A^2  \omega \gamma(k^2+\omega^2\gamma^{2}(1+C^2))}{(k^2+(1+C^2)\gamma^{2}\omega^2)^2-4k^2C^2 \gamma^{2} \omega^{2}}
\end{equation}

Similarly variance  $\langle V_{s} \rangle$ of the work averaged over a period of oscillation in is given by
\begin{equation}
\langle V_{s} \rangle = \langle W_{s}^{2} \rangle-\langle W_{s} \rangle^{2}
\end{equation}
\begin{equation}
\langle V_{s} \rangle =\frac{2}{\beta}\langle W_{s} \rangle
\end{equation}
In the time periodic state averaged input energy is dissipated into system as a heat[9].~Thus one can identify $\langle W_{s} \rangle$ as a hysteresis loss in the medium.~Since the  problem being linear we find that the time averaged hysteresis loss is independent of temperature.~However,~it depends  explicitly on the magnetic field and is a symmetric function of magnetic field.~Thus the magnetic field becomes a relevant variable in the time periodic asymptotic state.~However,~it must be noted that variance in the input energy can not be identified with the heat fluctuations[9].~In this time periodic state work done $W_{s}$ over a period  satisfies the SSFT,~ namely,~ equation(19).

It may be noted that the validity of SSFT for work done over a single period (equation(24)) is restricted only to overdamped linear models as in the present case.~In general this will not hold good in nonlinear situations[17].~~However,~~one can show that if one instead considers  work done over a large number of periods,~ indeed SSFT holds even for nonlinear models.~~The convergence of SSFT on accessible time scales have been studied in the previous literature[18]. ~~We have calculated separately the  work distribution for  $W_{s}$   in the inertial regime which satisfies SSFT.~~Moreover we have also shown that   one  can  obtain the orbital magnetic moment in this nonequilibrium state without violating the Bohr-van Leeuwen theorem.~~These results will be published elsewhere[16].

In conclusion we have solved analytically the work distribution of a charged particle in the presence of magnetic field in two different cases.~The first being the case where minimum of the harmonic potential is dragged with an uniform velocity and in the second case particle is subjected to an ac force.~For both cases JE is verified and this equality complements Bohr-van Leeuwen theorem on the absence of diamagnetism in classical system.~In case(i) we have shown that such a distribution in steady state does not depend on magnetic field and satisfies SSFT.~As opposed to this case in a time periodic assymptotic state for case(ii) magnetic field becomes a relevant variable.~The hysteresis loss over a cycle depends explicitly on the magnetic field.~Relaxation time in our system  can be controlled by magnetic field.~All our results are amenable to experimental verification  with charged beads in magnetic field.

\section{Appendix A:}
An expression  for $\langle W \rangle$  
\begin{widetext}
\begin{eqnarray}
\langle W \rangle&=&\frac{1}{\gamma[(k^2+\omega^2(1-C^2))^2+4 C^2 \omega 
^4]}\big[(k^{2}+\omega^{2}(1-C^{2}))\{\frac{-A^{2} k\sin^{2}
(\omega t)}{2}+\frac{A^{2} \omega^{2} t}{2}+\frac{A^{2} \omega \sin(2
\omega t)}{4}\}+2C\omega^2\{\frac{A^{2} C \omega^{2} t}{2}+\nonumber\\&&
\frac{A^{2} C \omega \sin(2\omega t)}{4}\}\big]+\frac{A^{2}
\omega^{2} \exp(-k^{*}t)}{2\gamma[(k^{2}+\omega^{2}(1-C^{2}))^{2}+4 C^{2} \omega^{4}]}\big[\frac{1}{k^{*^{2}}+(\Omega+\omega)^{2}}[\{(k^{2}
+\omega^{2}(1-C^{2}))(k^{*}+C(\Omega+\omega))+\nonumber\\&&2C\omega^{2}
(Ck^{*}-(\Omega+\omega))\}\cos(\Omega+\omega)t+\{2C\omega^{2}(k^{*}-C(\Omega+\omega))-(k^{2}+\omega^{2}(1-C^2))(\Omega+\omega+Ck^{*})\}\sin(\Omega+\omega)t]+\nonumber\\&&\frac{1}{k^{*^{2}}+(\Omega-\omega)^2}[\{(k^2+\omega^{2}(1-C^{2}))(k^{*}+C(\Omega-\omega))+2C\omega^{2}(Ck^{*}-(\Omega-\omega))\}\cos(\Omega-\omega)t+\nonumber\\&&\{2C\omega^2(k^{*}-C(\Omega-\omega))-(k^{2}+\omega^{2}(1-C^{2}))(\Omega-\omega+Ck^{*})\}\sin(\Omega-\omega)t]\big]+c^{'} ,
\end{eqnarray}
\end{widetext}
where
\begin{eqnarray}
c^{'}=\frac{a_{1}a_{2}}{c_{1}c_{2}},
\end{eqnarray}
$a_{1}=-2A^{2}\omega^{2}(k^{*^{2}}+\Omega^{2}+\omega^{2})$,\\
$a_{2}=k^{*}\{k^{2}+\omega^{2}(1+C^{2})\}+\Omega C(k^{2}+3\omega^{2}-\omega^{2}C^{2})$,\\
$c_{1}=(k^{*^{2}}+\omega^{2}(1-C^2))^2+4 C^2 \omega ^4$,\\
$c_{2}=2[k^{*^{2}}+(\Omega-\omega)^2][k^{*^{2}}+(\Omega+\omega)^2]$.\\

 Here $\Omega=\frac{k C}{1+C^2}$, $k^{*}=\frac{k}{1+C^{2}}$ and $C=\frac{\vert e \vert B }{\gamma c}$ with $k$ here written  for $\frac{k}
{\gamma}$.\\
In the absence of magnetic field,~ $\langle W \rangle$ reduces to
\begin{eqnarray}
\langle W\rangle&=&-\frac{A^2 k}{2\gamma(k^2+\omega^2)}{\sin^2(\omega t)}
+\frac{A^2\omega}{4\gamma(\omega^2+k^2)} \sin(2\omega t)+\nonumber\\ &&
+\frac{A^2 \omega^2 t}{2\gamma(k^2+\omega^2)}- \frac{A^2\omega^2k}
{\gamma(k^2+\omega^2)^2}+\frac{A^2\omega^2}{\gamma(k^2+\omega^2)^2}
\nonumber\\ &&\exp(-kt)[k\cos(\omega t)-\omega \sin(\omega t)] 
\end{eqnarray}
\section{Acknowledgements}
One of us (AMJ) thanks Abhishek Dhar for helpful discussions.


\begin{thebibliography}{999}
\bibitem{Bustamante} C.~Bustamante,~J.~Liphardt  and  F.~Ritort,   Physics Today  {\bf 58},  45   (2005).
\bibitem{Evans}  D.~J.~Evans  and  D.~J.~Searls,   Adv.Phys.  {\bf 51}, 1529  (2002).
\bibitem{Jarzynski} C.~Jarzynski, Phys.Rev. Lett. {\bf 78}, 2690 (1997);~~~Phys.~Rev.~ E  {\bf 56}, 5018  (1997).
\bibitem{Crooks} G.~E.~Crooks,  Phys.Rev. E  {\bf 60}, 2721  (1999);~~~Phys.Rev. E  {\bf 61}, 2361 (2000).
\bibitem{Hatano}  F.~Hatano  and  S.~Sasa, Phys.Rev. Lett. {\bf 86}, 3463 
(2000).
\bibitem{Liphardt}  J.~Liphardt  et.al,    Science. {\bf 296}, 1833(2002);~~~F.~Douarche   et al,    Europhys. Lett. {\bf 70}, 593 (2005);~~~G.~M.~Wang   et al,    Phys. Rev. Lett. {\bf 89}, 050601 (2002);~~~O.~Narayan and A.~Dhar,   J. Phys.A:Math Gen {\bf 37}, 
63 (2004).
\bibitem{Trepangnier}  E.~M.~Trepangnier  et al,  Proc. Natt. Acad. Sci. {\bf 101}, 15038 (2004).
\bibitem{Lechner} W.~Lechner  et.al,  J. Chem. Phys. {\bf 124}, 044113 (2006);~~~G.~Hummer  and  A.~Szabo,  Proc. Natl. Acad. Sci. {\bf 98}, 3658 (2001).
\bibitem{Zon} R.~van~Zon  and E.~G.~D.~Cohen,  Phys.Rev. E  {\bf 67}, 046102 (2002);~~~Phys.Rev. E   {\bf 69}, 056121  (2004).
\bibitem{Mazonka}  O.~Mazonka   and C.~Jarzynski,  Cond-mat/ 9912121.
\bibitem{Ritort}  F.~Ritort,   J.~Stat. Mech.  P10016 (2004);~~~R.~C.~Lua   
and A.~Y.~Grosberg,  J.Phy.Chem.B  {\bf 109}, 6805 (2005);~~~L.~Bena,~C.~ 
Van ~den ~Broeck  and R.~ Kawai,  Europhys.Lett.  {\bf 71}, 879  
(2005);~~~A.~Dhar,   Phys. Rev. E  {\bf 71}, 036126 
(2005);~~~R.~Marathe  and A.~Dhar,   Phys. Rev. E  {\bf 72}, 066112 (2005);~~~A.~Imparato   and  L.~Peliti,   Europhys.Lett.  {\bf 69}, 643 (2005).
\bibitem{chandrasekhar}   S.~Chandrasekhar,  Rev. Mod. Phys.   {\bf 15}, 1 (1943).
\bibitem{Jayannavar}     A.~M.~Jayannavar   and N.~Kumar,  J. Phys. A  {\bf 14}, 1399  (1981).
\bibitem{Bohr}     N.~Bohr, Dissertation,   Copenhagen  (1911);~~~J.~H.~van Leeuwen,  J. Phys.  {\bf 2}, 361  9(1921);~~~R.~E.~Peierls,    Surprises in theoritical Physics  (Princeton University Press,  Princeton, (1979)).
\bibitem{Ris} H. Risken, {\em {The Fokker Planck Equation}},~
    Springer Verlag, Berlin, 1984.
\bibitem{Jayannavar}   A.~M.~Jayannavar    and  M.~Sahoo  [to be published].
\bibitem{Saikia}  S.~Saikia, R.~Roy  and  A.~M.~Jayannavar, 
cond-mat/0701303.
\bibitem{Reid}   J.~C.~Reid,~  et.~al,   Phys. Rev. E  {\bf 70},~016111 (2004);~~~ J.~C.~Reid, E.~M.~Sevick and  D.~J.~Evans,  Europhys.~Lett.  {\bf 5},~726 (2005);~~~G.~M.~Wang  et.al,  Condens.~Matter  {\bf 17},
~s3239  (2005);~~~F.~Douarche,~  et.~al,  Phys. Rev. Lett.  {\bf 97},~ 140603   (2006).
\end{thebibliography}
\end{document}